\documentclass[reprint,aps,pre,onecolumn,notitlepage,superscriptaddress]{revtex4-1}
\usepackage{graphicx}
\usepackage{amsmath,amssymb}

\usepackage{graphicx}
\usepackage{dcolumn}
\usepackage{bm}
\usepackage{comment}

\newcommand{\kbend}{\kappa_\mathrm{bend}/\kappa_\mathrm{LJ}}
\newcommand{\kfbend}{\kappa_\mathrm{bend}/\kappa_\mathrm{FENE}}
\usepackage{color}

\begin{document}

\title{Interpretation of the vibrational spectra of glassy polymers using coarse-grained simulations}

\author{Rico Milkus}
\author{Christopher Ness}
\author{Vladimir V. Palyulin}
\author{Jana Weber}
\author{Alexei Lapkin}
\author{Alessio Zaccone}
\affiliation{Department of Chemical Engineering and Biotechnology, University of Cambridge, Cambridge CB3 0AS, United Kingdom}

\date{\today}

\begin{abstract}
The structure and vibrational density of states (VDOS) of polymer glasses are investigated using numerical simulations based on the classical Kremer-Grest bead-spring model.
We focus on the roles of chain length and bending stiffness, the latter being set by imposing three-body angular potentials along chain backbones.
Upon increasing the chain length and bending stiffness, structural reorganisation leads to volumetric expansion of the material and build-up of internal stresses.
The VDOS has two dominant bands: a low frequency one corresponding to inter- and intra-chain non-bonding interactions
and a high frequency one corresponding principally to vibrations of bonded beads that constitute skeletal chain backbones.
Upon increasing the steepness of the angular potential, vibrational modes associated with chain bending gradually move from the low-frequency to the high-frequency band. This redistribution of modes is reflected in a reduction of the so-called Boson peak upon increasing chain stiffness. Remarkably, the finer structure and the peaks of the high-frequency band, and their variations with stiffness, can, for short chains, be explained using an analytical solution derived for a model triatomic molecule. For longer chains, the qualitative evolution of the VDOS with chain stiffness is similar, although the distinct peaks observed for short chains become increasingly smoothed-out. Our findings can be used to guide a systematic approach to interpretation of Brillouin and Raman scattering spectra of glassy polymers in future work, with applications in polymer processing diagnostics. 
\end{abstract}
\maketitle


\section{Introduction}

Raman spectroscopy can detect vibrational and electronic properties of materials
over a broad range of temperatures and pressures,
and is a well-established and widely-used non-destructive measurement technique~\cite{zhao2004raman,grasselli1993industrial}.   
Comprehensive predictive models for Raman and Brillouin spectra 
are important for many applications involving amorphous carbon-based materials,
from nanotechnology to polymer reaction engineering~\cite{houben2015feasibility,frauendorfer2010polymerization,hergeth2003industrial,ferrari2006raman}.
Of particular interest is emulsion polymerisation~\cite{elizalde2005monitoring},
a common manufacturing route for many rubbers and plastics.
The complexity of this process hinders characterisation of product quality by traditional methods~\cite{kiparissides1996polymerization}, 
and it is increasingly being probed by Raman spectroscopy.

The vibrational density of states (VDOS) of solids is the main input for
the prediction of the Raman and Brillouin scattering spectra.
For glasses, the Shuker-Gammon formula gives the Raman intensity as a function of the VDOS as~\cite{shuker1970raman}
\begin{equation}
I(\omega)=\frac{n(\omega)+1}{\omega}C(\omega) D(\omega) \text{,}
\end{equation}
where $n(\omega)+1$ is the Bose-Einstein occupation factor,
$D(\omega)$ is the VDOS,
and $C(\omega)$ is the photon-phonon coupling coefficient.
Since $C(\omega)$ is a simple function of frequency, possibly quadratic~\cite{martin1974model},
it is clear that most of the structure of the Raman spectrum is directly related to the $D(\omega)$ spectrum.
While the VDOS of crystals can be obtained by a straightforward exercise in Fourier analysis,
the same problem for amorphous solids, such as glasses, is analytically intractable and presents a rich phenomenology.
This phenomenology is yet more complex when the building blocks are polymer chains, which,
in the disordered glassy state, can have a considerably larger variety of conformations. 

There have been numerous studies into the vibrational properties of polymeric systems~\cite{herzberg1945molecular,bower1992vibrational},
starting from theoretical determinations of the single-chain backbone vibrational spectra in seminal works by~\citet{kirkwood1939skeletal} and~\citet{pitzer},
followed by the powerful combination of Wilson's GF-method with group theory by~\citet{higgs1953vibration}.
These methods are not applicable to polymer glasses, however,
where the chain conformation does not possess any periodicity that can allow the application of group theoretical methods.
Further advances in numerical techniques have focused on reducing the computational time of the diagonalization problem~\cite{zerbi1971vibrational}. 

While signatures of individual monomers and their constituent bonds are very well characterized in the vibrational bands of highest energy in the spectrum,
the relation between coarse-grained polymer structures and vibrational properties in the low frequency part of the spectrum is relatively unexplored.
In the contemporary literature, the use of coarse-grained systems as model materials for studying the vibrational properties of amorphous solids
has become a standard approach~\cite{damart2017theory,ilyin2009randomness,lemaitre2006sum,zaccone2011approximate}.
In this direction, coarse-grained simulations based on the classical Kremer-Grest model~\cite{kremer1986dynamics}
for bead-spring polymers can enable direct calculation of the VDOS.
With a suitable procedure for coupling the VDOS to the Raman spectra~\cite{ilyin2009randomness},
such numerical approaches will be able to offer a systematic approach to linking vibrational properties to coarse-grained structures for polymers of arbitrary length and monomer-monomer interactions.

Here, we report the structural properties and the VDOS for coarse-grained polymer glasses as functions of the chain length and the chain bending stiffness.
We identify clear trends in the vibrational spectra that derive from microstructural rearrangements
as the chain length and chain stiffness increase.
Through these quantities, it will be possible in future work to make predictions about how the experimentally observed vibrational spectra will evolve during the course of an emulsion polymerisation, for example,
guiding the development of noninvasive industrial process monitoring techniques.
This work can further serve as the basis for quantitative understanding and modelling of Raman and Brillouin spectra at the atomistic level,
particularly by coupling to atomistic simulation techniques.

In the following, we first describe the numerical method used, then go on to study the structural and volumetric changes as functions of varying chain length and stiffness.
We then analyze the VDOS as a function of chain length and stiffness, providing a mechanistic interpretation informed by an analytical argument.

\section{Simulation details}
Our model uses a coarse-graining approach that treats
polymer chains as linear series of monomer `beads' on an elastic string.
In a harmonic approximation, monomeric scale physics dominate the region of the VDOS of interest to this work,
and indeed govern the viscoelastic response of the material~\cite{palyulin2017instantaneous}.
For each bead in the system we use \texttt{LAMMPS}~\cite{plimpton1995fast} to solve the Langevin equation
\begin{equation}
m\frac{d{v}}{dt} = -\frac{m}{\xi}{v} - \frac{d{U}}{dr} + {f}_B(t) \text{,}
\end{equation}
for uniform beads of mass $m$ and velocity ${v}$, coefficient of friction $m/\xi$ and random forces ${f}_B(t)$ satisfying $\langle f_B(t)f_B(t')\rangle = 2mk_BT\delta(t-t')/\xi$.
Beads interact with each other through a potential $U$, given by the Kremer-Grest model~\cite{kremer1986dynamics}
with the addition of angular potentials that impose bending constraints on triplets of three consecutive beads along the chain backbones.
Overall, the model for the potential energy $U$ comprises three terms:
        \begin{figure}
        \includegraphics[trim={102mm 115mm 0mm 0mm},clip,width = 0.275\textwidth]{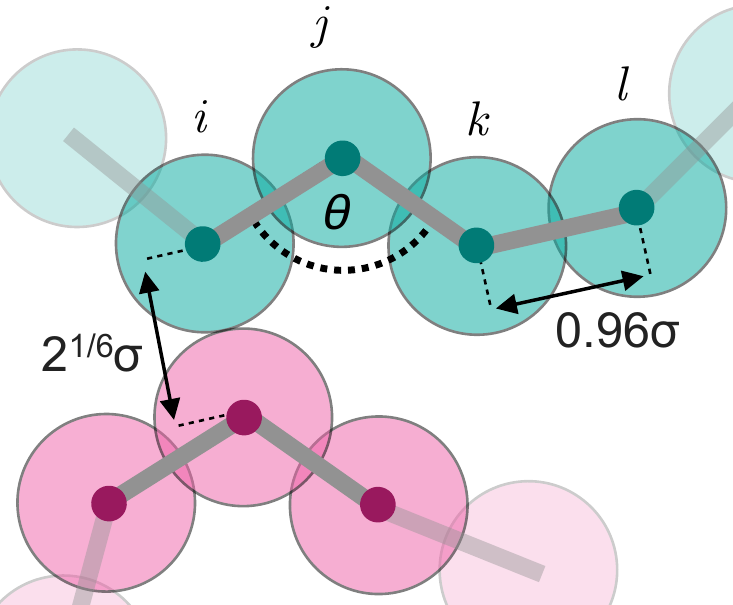}
        \caption{
        Sketch of simulated polymer system showing two interacting chains (blue and pink). Shown are the LJ rest position of $2^{1/6}\sigma$ between non-bonded beads, the FENE rest position of $0.96\sigma$ between beads along a single linear chain, and the angle $\theta$ formed between three consecutive beads (forming a triplet) along chains.
        }
        \label{fig:sketch}
        \end{figure}
[i] A truncated and shifted Lennard-Jones (LJ) potential of form
\begin{equation}
U_\mathrm{LJ}(r) = 4\varepsilon_\mathrm{LJ} \left[\left(\frac{\sigma}{r}\right)^{12} - \left(\frac{\sigma}{r}\right)^6 - \left(\left(\frac{\sigma}{r_c}\right)^{12} - \left(\frac{\sigma}{r_c}\right)^6\right) \right] \text{,}
\end{equation}
acting between all bead pairs within a cut-off range $r_c=2.5\sigma$,
where $r$ is the bead-bead separation,
$\sigma$ is the zero-crossing distance for the potential and the prefactor $\varepsilon_\mathrm{LJ}$ sets the LJ energy scale.
Setting $\frac{dU_\mathrm{LJ}(r)}{dr} = 0$ leads to an energy minimum and corresponding LJ rest position at $2^{1/6}\sigma$.
The LJ potential effectively acts as an excluded volume, as illustrated in Figure~\ref{fig:sketch};
[ii] A finitely extensible nonlinear elastic (FENE) potential acting between sequential bead pairs along each linear chain
\begin{equation}
U_\mathrm{FENE}(r) = -0.5\varepsilon_\mathrm{FENE} R_0^2 \ln \left[1 - \left(\frac{r}{R_0}\right)^2\right] \text{,}
\end{equation}
where $R_0$ is the maximum FENE bond length and $\varepsilon_\text{FENE}$ is the bonding energy scale.
Adjacent beads along polymer chains have an overall interaction that represents the sum of the Lennard-Jones and FENE potentials,
giving a rest position for bonded beads (obtained by setting $\frac{d}{dr}(U_\mathrm{LJ}(r)+U_\mathrm{FENE}(r))=0$)
as $\approx0.96\sigma$ for the parameters used throughout this work.
This discrepancy relative to the LJ rest length gives sufficient bidispersity to suppress crystallisation throughout,
and we set $\varepsilon_\mathrm{FENE}/\varepsilon_\mathrm{LJ} = 30$;
[iii] An energy associated with chain bending, given by
\begin{equation}
U_{\mathrm{bend}}(\theta) = \varepsilon_\mathrm{bend}[1-\cos(\theta-\theta_0)]
\end{equation}
for energy scale $\varepsilon_\mathrm{bend}$.
The angle $\theta$ is formed between three consecutive beads (a triplet) along the length of the linear chains (Figure~\ref{fig:sketch}),
and the characteristic rest angle is $\theta_0 = 109.5^\circ$.
The resistance to bending of the polymer chains is thus set by $\varepsilon_\mathrm{bend}$,
which is related to the persistence length $l_{p}$ of the chain via the standard relation: $l_{p}=\varepsilon_\mathrm{bend} \sigma /k_b T$.

The relative importance of the three potentials in setting the
overall structure and dynamics of the polymers is determined by their prefactors
$\varepsilon_\mathrm{LJ}$, $\varepsilon_\mathrm{FENE}$ and $\varepsilon_\mathrm{bend}$.
Since each of the potentials has a different form, it is difficult to compare the values of these prefactors directly.
In order to render the different interaction strengths more comparable, therefore,
we find it convenient to take a harmonic approximation about the rest position of each potential and consider the resulting spring constants $\kappa$.
We find $\kappa_\mathrm{LJ} \approx 57.1\varepsilon_\mathrm{LJ}/\sigma^2$, $\kappa_\mathrm{FENE} \approx 32.7\varepsilon_\mathrm{FENE}/\sigma^2$
and $\kappa_\mathrm{bend} = \varepsilon_\mathrm{bend}/\sigma^2$.
To characterize our systems we use two control parameters $\kbend$ and $\kfbend$ that compare the bending stiffness to the LJ and FENE bond strength, respectively.
The strength of FENE bonds is fixed such that $\kappa_\mathrm{FENE}/\kappa_\mathrm{LJ} \approx 17.2$ throughout (recalling that $\varepsilon_\mathrm{FENE}/\varepsilon_\mathrm{LJ} = 30$).
We explore bending stiffnesses in the range $\kbend = 0 \to 20$.
A sketch of two interacting polymer chains is shown in Figure~\ref{fig:sketch},
highlighting the angle $\theta$ on which $U_\mathrm{bend}$ acts as well as the rest positions for LJ ($2^{1/6}\sigma$) and FENE ($0.96\sigma$) interactions.

With reference to fundamental units of mass $\mu$, length $d$, and energy $\epsilon$, we set $\sigma=1$, $R_0=1.5$, $m=1$ and $\varepsilon_\mathrm{LJ}=1$,
giving a time unit of $\tau=\sqrt{m\sigma^2/\varepsilon_\mathrm{LJ}}$, and we set $\xi=100\tau$.
The system volume $V$ has units $d^3$.
A dissipative timescale emerges as $m\sigma^2/\xi\varepsilon_\mathrm{LJ}$, and a thermal timescale emerges as $m\sigma^2/\xi k_BT$ (where $k_B := 1$ [units energy/temperature]).
The state of our system, i.e. whether it is in the melt or glassy state, is simply given by the ratio of these timescales, as $T^*=k_BT/\varepsilon_\mathrm{LJ}$.
Two additional rescaled temperatures could be defined using $\varepsilon_\mathrm{FENE}$ or $\varepsilon_\mathrm{bend}$ as the reference energy,
but we find that the most convenient description and characterisation of the transition to glassy behavior is obtained using $\varepsilon_\mathrm{LJ}$.

Initial loose polymer configurations are generated within a cubic periodic domain using a non-overlapping random-walk algorithm.
We use a system of $N_p = 5\times10^3$ beads, in chains of uniform length $L$, which we vary from 2 to 50. The value of $N_p$ is chosen following the entanglement critical of Ref~\cite{sliozberg2012bead}, and moreover we demonstrate the sensitivity to $N_p$ in Figure~\ref{fig:dos_split}.
For each value of $L$ we generate 5 realisations of the system for the purposes of ensemble averaging.
We comment on the variation between realisations elsewhere~\cite{palyulin2017instantaneous}.
The system is first equilibrated in a melted state at $T^*=1.2$, maintaining zero external pressure using a Nose-Hoover barostat with damping parameter of $100\tau$.
The system is subsequently cooled to $T^*=0.1$ by decreasing $T^*$ at rate $1/\tau_c$, with $\tau_c \sim \mathcal{O}(10^5)\tau$. 
Since $T^*=0.1$ is below the glass transition for all of the polymers considered in this work,
this cooling procedure allows us to measure $T_g$.
For determining the vibrational properties, though, it is necessary to go to lower temperatures.
To reach temperatures closer to $T^*=0$,
we subsequently relax the system further by applying the gradient method to the simulation configuration at constant volume.
We used the fluctuation of net forces acting on beads $\langle f_i^2\rangle \sim T^*$ as a measure for temperature. By comparing the forces with the reference value from $T^*=0.1$ we get the temperature of the relaxed configuration by $T^* = 0.1 \langle f_i^2\rangle/\langle f_i^2(T^*=0.1)\rangle$.
By this protocol, a target temperature of $T^*=10^{-4}$ was reached for each realization. Further decreasing the temperature does not lead to changes in the VDOS or structural quantities.

\section{Structure of coarse-grained polymer glasses}

\subsection{Changes in $T_g$ with chain length and stiffness}

        \begin{figure}
        \includegraphics[trim={0 0 80mm 0},clip,width = 0.48\textwidth]{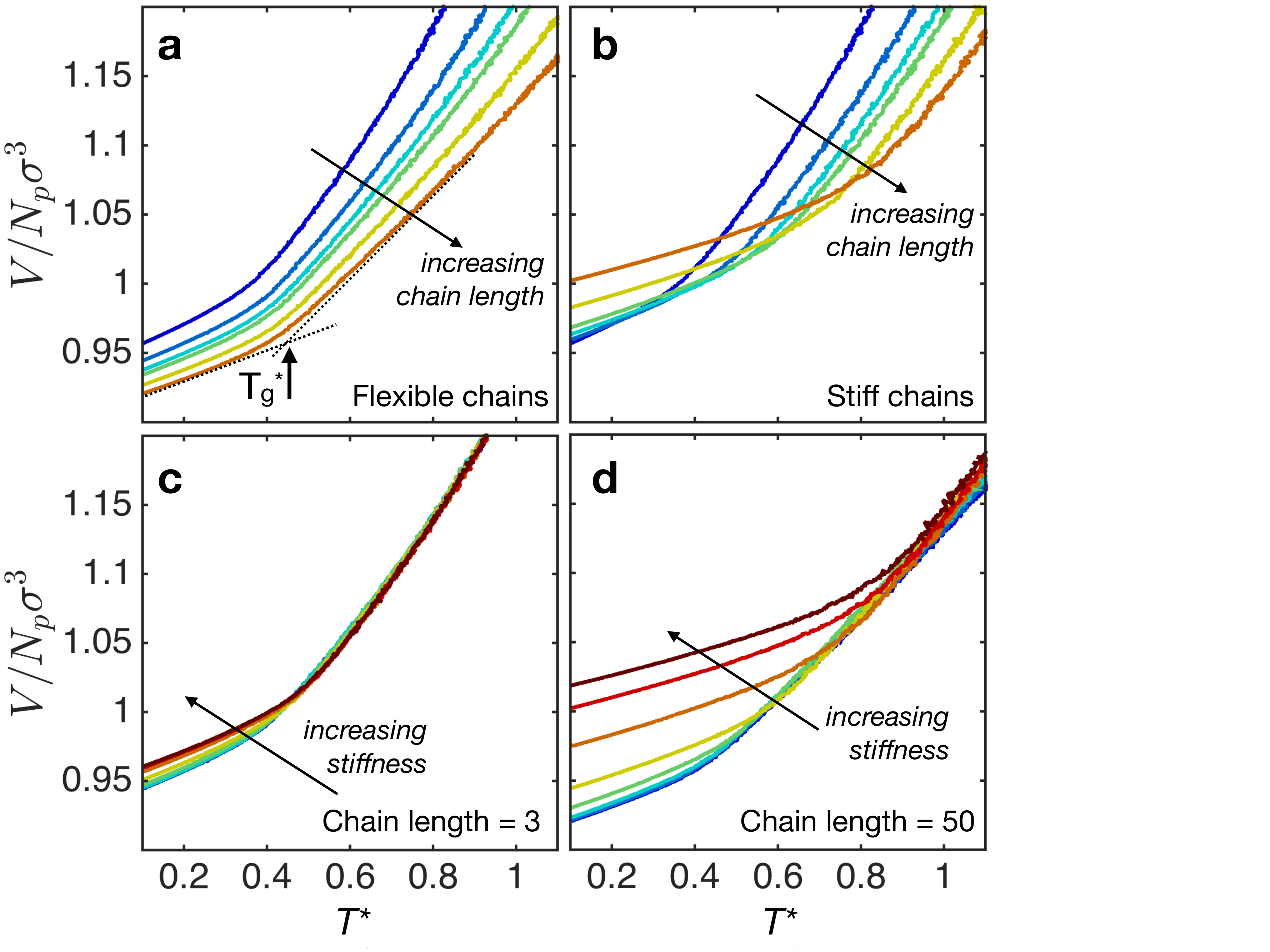}
        \caption{Volume-temperature curves for polymer cooling as functions of chain length and chain stiffness. The volume is rescaled with bead number $N_p$ and size $\sigma^3$.
        a) Increasing chain length for $\kbend=0$. Highlighted is the glass transition temperature $T_g^*$, where the polymer transitions from a melt to a glassy state;
        b) Increasing chain length for $\kbend=17.5$;
        c) Increasing chain stiffness for chain length 3;
        d) Increasing chain stiffness for chain length 50.
        Colors in (a) and (b) refer, from blue to red, to chain lengths 2, 3, 4, 5, 20 and 50.
        Colors in (c) and (d) refer, from blue to red, to chain stiffnesses $\kbend$ = 0.0175, 0.0525, 0.175, 0.525, 1.75, 5.25 and 17.5.
        }
        \label{fig:tg}
        \end{figure}
        
Ensuring that the external pressure remains zero,
the system undergoes a decrease in volume $V$ as it is cooled, Figure~\ref{fig:tg}.
In Figure~\ref{fig:tg}a, a change of gradient is identified at $T^*=T_g^*$,
corresponding to the glass transition~\cite{han1994glass,white2016polymer}.
As reported in Figures~\ref{fig:tg}a-b,
the model predicts that $T_g^*$ increases with the chain length~\cite{boyer1974variation},
consistent with the classical free-volume result of Flory and Fox~\cite{fox1950second},
and with the more recent criterion based on generalized Born melting for glasses~\cite{zaccone2013disorder}.
This is the case for both fully flexible ($\kbend=0$) and very stiff ($\kbend=17.5$) chains.
As expected~\cite{strobl1997physics}, we further find that $T_g^*$ increases with $\kbend$ (Figure~\ref{fig:tg}c-d),
with apparent limiting values occurring for $\kbend\to0$ and $\kbend > 4$.
The increase is significantly more pronounced as $L$ is increased.
We reported the increase of $T_g^*$ with $\kbend$ and provide further details in our earlier article~\cite{ness2017nonmonotonic}.
        
We find that $T_g^*$ varies between $\approx 0.4$ and $\approx 0.9$ for all values of $L$ and $\kbend$,
and that in all cases the system is well within the glassy state at $T^*=0.1$.
When we increase $\kbend$ above 20 (corresponding to $\kappa_\mathrm{bend}/\kappa_\mathrm{FENE}\approx 1$),
we find that the angular potentials are large enough to stretch the FENE bonds beyond their maximum length $R_0$
at which point the chains break and the simulation becomes unstable due to the divergence of the $\ln(1-\left(r/R_0\right)^2)$ term in the FENE expression.
We therefore treat this as a limiting value of $\kbend$ and do not explore stiffer chains.

\subsection{Changes in density with chain length and chain stiffness}

        \begin{figure}
        \includegraphics[trim={0 0 0mm 0},clip,width = 0.45\textwidth]{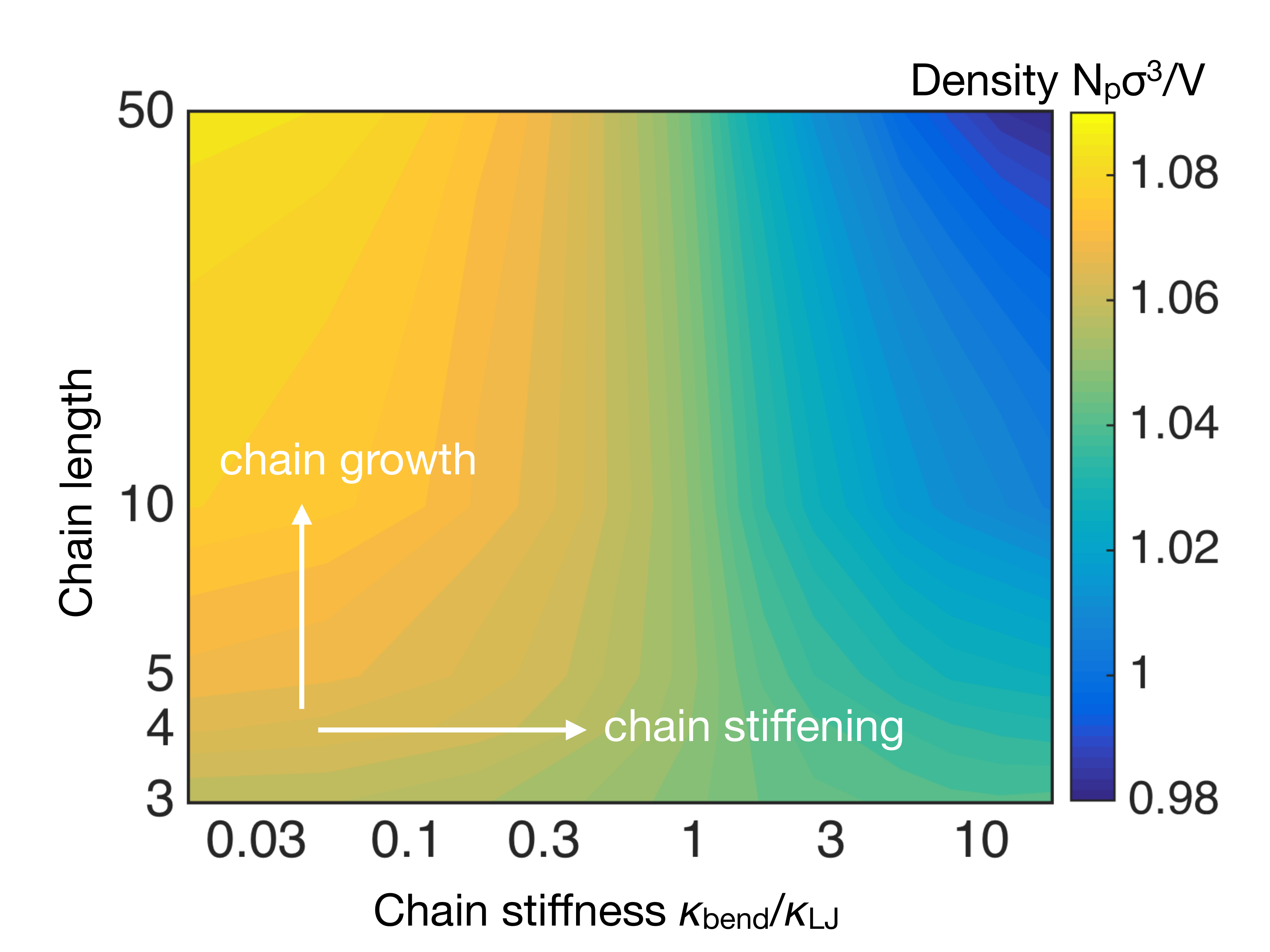}
        \caption{Contour plot of polymer glass density (enumerated as $N_p\sigma^3/V$ at $k_bT/\varepsilon_\text{LJ}=0.1$) as functions of chain length and chain stiffness. We note that increasing chain length leads to compaction for flexible chains but expansion for stiff chains.}
        \label{fig:density}
        \end{figure}

It is evident from Figure~\ref{fig:tg} that the value of $V/N_p\sigma^3$ at $T^*= 0.1$
is sensitive to both chain stiffness and length.
In our earlier article~\cite{ness2017nonmonotonic},
we showed that differences in $V/N_p\sigma^3$ persist even at fixed $T^*/T_g^*$ (rather than fixed $T^*$),
demonstrating that there are robust changes in density as chain bending and length are varied. 
We present in Figure~\ref{fig:density} a contour map of polymer glass density,
quantified as the number of beads per unit volume $N_p/V$,
rescaled by the characteristic bead excluded volume $\sigma^3$,
measured at $T^*=0.1$. Since our subsequent minimization protocol conserves volume,
the map also applies at $T^*=10^{-4}$.
With respect to the density at $L=3$,
it is interesting to note that increasing chain length leads to compaction for flexible chains but expansion for stiff chains.

For very flexible chains ($\kbend\ll1$),
the key effect of increasing chain length is to move bead pairs from the LJ rest position at
$2^{1/6}\sigma$ to the FENE rest position at $0.96\sigma$,
while maintaining a purely central-force system with minimal explicit bending constraints.
As a result, the density increases with increasing chain length as illustrated in Figure~\ref{fig:density}
and similarly by the decreasing value of $V/N_p\sigma^3$ at $T^*=0.1$ in~Figure~\ref{fig:tg}a.

For less flexible chains, the roles of stiffness and chain length are more subtle.
In order to achieve mechanical stability at, and below, $T_g^*$,
approximately monodisperse beads in central force networks
require six pairwise interactions to fully constrain their translational degrees of freedom,
in agreement with Maxwell's criterion for isostaticity.
As bending stiffness is increased,
the translational motions of beads along chain backbones become
increasingly constrained by three(and many)-body interactions.
This means that as $\kbend$ increases,
the translational degrees of freedom
of individual beads can be fully constrained
with fewer than six pairwise interactions per bead~\cite{zaccone2013disorder,hoy2017jamming}.
We quantified this effect using the coordination number $Z$ in another contribution~\cite{ness2017nonmonotonic}.
Since we operate at fixed external pressure,
this lower coordination further implies that marginal stability can be achieved
at lower density as stiffness is increased.
This is the result observed in Figure~\ref{fig:density} for $L>3$ and indeed in
Figure~\ref{fig:tg}c-d (inverse density $V/N_p\sigma^3$ at $T^*=0.1$ increases with increasing $\kbend$),
with the effect becoming more evident for longer chains, which permit many-body interactions.

As we deviate from the short chain limit,
it is interesting to note that there is a continuous transition from very weak dependence
to rather strong dependence on stiffness with increasing chain length.
This can be interpreted by considering again the coordination argument above,
which argued that bending constraints impose many-body effects along chains
such that marginal stability can be achieved with fewer \emph{pairwise} contacts than otherwise.
For shorter chains, the maximum number of beads than can be correlated with one another in this way is small,
so many-body interactions only have a weak contribution to the overall stability of the material.
As such, when the stiffness is increased in short chains where there aren't many angular potentials defined (in relative terms),
most of the interactions remain as central force and as such the density varies only weakly.
Conversely, for long chains,
the increased stiffness allows many-body bending constraints to affect a
higher proportion of the overall number of interactions, so the density decrease becomes exaggerated.
Interestingly, at $L=50$, the variation of density with chain stiffness is not linear,
but rather it has an inflection around $\kbend\approx 1$.

\subsection{Deviations from rest positions and the resulting internal stresses}

        \begin{figure}
        \includegraphics[trim={0 18mm 0mm 0},clip,width = 0.475\textwidth]{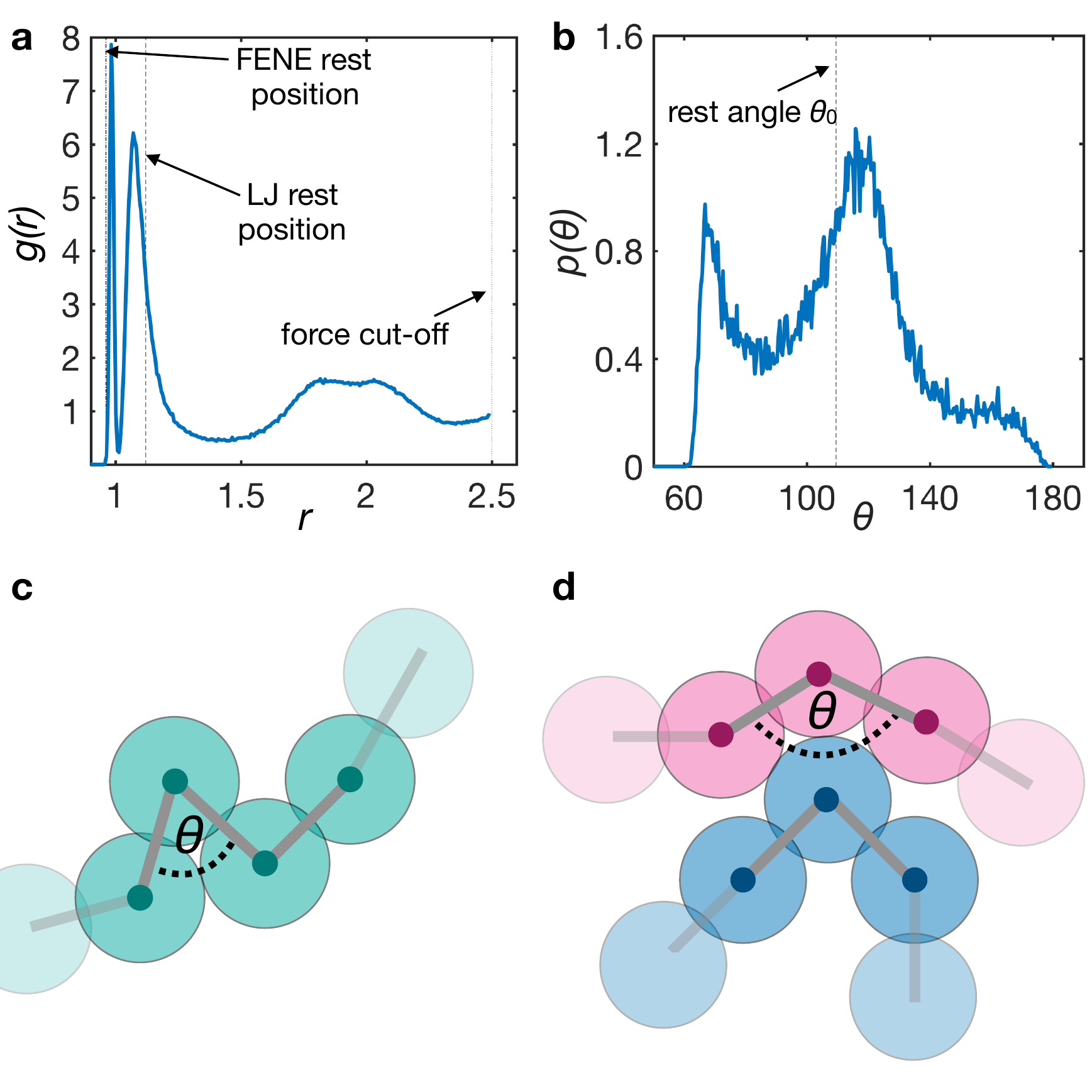}
        \caption{(a) Radial distribution function $g(r)$ and (b) angular distribution function $p(\theta)$ for chains with $L=10$ and $\kbend=0.175$. Shown are the FENE rest position, LJ rest position and the resting angle, as well as the force cut-off beyond which we do not compute LJ interactions. (c) and (d) illustrate the implicit angular resting positions that arise due to two different configurations of the excluded volumes of beads.}
        \label{fig:gr}
        \end{figure}

                        \begin{figure*}
        \includegraphics[trim={0 0 0mm 0},clip,width = 0.8\textwidth]{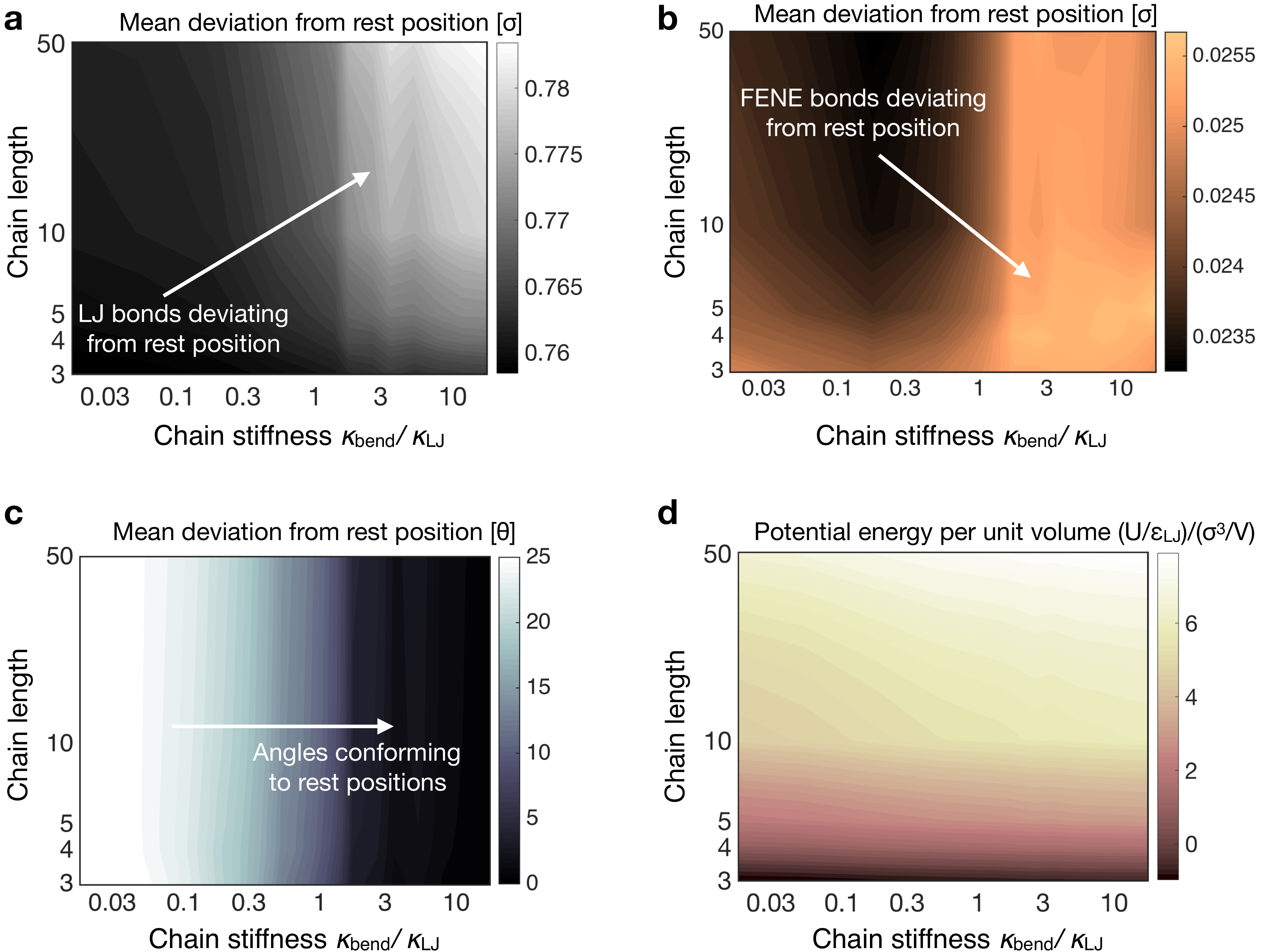}
        \caption{
        Structural origin of internal stresses in the polymer glasses as a function of chain length and stiffness. Shown are
        a) Mean deviation of LJ pairs from their rest position;
        b) Mean deviation of FENE bonds from their rest position;
        c) Mean deviation of angles from their rest position, showing approximate independence of chain length;
        d) Total potential energy per unit volume, equivalent to the internal stress of the material.
        }
        \label{fig:maps}
        \end{figure*}

It is likely that steric constraints will play a role within our densely packed systems,
meaning that beads will not typically be situated at their minima with respect to all three potentials (LJ, FENE, bending)
even when the temperature is considerably less than $T_g^*$.
Such deviations from minima will lead naturally to internal stresses in the material.
As chain length and bending stiffness are increased,
it is likely that the extent to which beads deviate from their respective rest positions,
and hence the total internal stress in the material, will change.
It was shown recently~\cite{lin2016evidence} that properly accounting
for internal stresses in jammed emulsions is crucial to correctly obtaining the VDOS.
Here we give a description of the source of internal stresses,
which will help to guide our interpretation of the VDOS below.

To illustrate the deviation of beads from their rest positions,
we present in Figure~\ref{fig:gr} an example bead-bead radial distribution function $g(r)$
and angular distribution function $p(\theta)$
for chains with $L=10$ and $\kbend=0.175$ in the minimized configuration,
highlighting the specified rest positions for LJ, FENE and bending potentials as well as the LJ cut-off distance.
There is a clear deviation from each of the rest positions.
In particular, LJ-bonded beads lie,
on average, closer than their rest positions dictate,
while FENE bonds are typically stretched, Figure~\ref{fig:gr}a.
Although the FENE bonds have a steeper potential than their LJ counterparts,
the LJ bonds are far more widespread in the system,
being both longer range and inter-chain.
Thus it is likely that the mean positions of FENE-bonded pairs
can be dictated by the LJ bonds in their immediate vicinity.
Deviations of both LJ and FENE naturally lead \textit{locally} to internal stresses
and thus to a net storage of potential energy in the system.

Furthermore, Figure~\ref{fig:gr}b shows peaks in the angular distribution at around
$\theta = 70^\circ$ and $\theta=120^\circ$
that are not related to the potential minimum at $\theta_0 = 109.5^\circ$.
Rather, implicit angular constraints that generate energetically favorable configurations
arise in the material due to the LJ and FENE potentials.
We illustrate such configurations in Figures~\ref{fig:gr}c-d.
These configurations nonetheless
lie far from their bending potential minima,
thus accumulating additional potential energy in the system.
The broad $p(\theta)$ distribution steadily narrows
as the value of $\kbend$ is increased, as we showed in an earlier article~\cite{ness2017nonmonotonic}.

To quantify the potential energy in the system, we compute
the mean displacement of beads from their rest positions as a function of chain length and stiffness.
At any time, there are $N_\mathrm{LJ}$ Lennard-Jones interactions
(including those up to the cut-off $r_c=2.5\sigma$),
which we label with the index $\alpha$ (so the bead-bead distance is  $r_\alpha$).
We take the average magnitude of the deviation of $r_\alpha$ from the LJ rest position.
Analogous calculations are done for FENE interactions and bending interactions.
Overall, we compute:
$
\frac{1}{N_\mathrm{LJ}}\sum_\alpha|r_\mathrm{\alpha}-2^{1/6}\sigma|
$, $
\frac{1}{N_\mathrm{FENE}}\sum_\beta|r_\mathrm{\beta}-0.96\sigma|
$ and $
\frac{1}{N_\mathrm{bend}}\sum_\gamma|\theta_\mathrm{\gamma}-\theta_0|
$.
The results are given in Figure~\ref{fig:maps}a-c for LJ, FENE and bending interactions respectively.

As bending stiffness $\kbend$ increases,
the chains increasingly conform to $\theta_0$ independently of chain length (Figure~\ref{fig:maps}c).
This requires changes in structure that must be accommodated by small additional deviations of LJ and FENE bonds from their resting positions.
For LJ interactions (Figure~\ref{fig:maps}a) the deviations increase steadily with stiffness.
This effect is more marked for long chains,
which have a larger number of bending constraints per bead.
For FENE bonds (Figure~\ref{fig:maps}b), meanwhile, the behavior is more subtle.
The primary effect is similar to that observed for LJ interactions:
increasing bending stiffness increases the deviation of FENE bonds from their rest positions.
For short and flexible chains, however, there is an anomalous secondary effect:
for $\kbend<1$, the deviation of FENE bonds increases with \emph{decreasing} chain length. 
As discussed above, the ubiquity of LJ interactions means they may dictate
the positions of FENE bonded beads regardless of their considerably weaker potential.
As chain length decreases, the average number of FENE bonds
per bead decreases steadily, while the number of LJ interactions per bead remains unchanged.
This means that LJ interactions can have more influence on FENE positions as chain length decreases,
thus leading to additional stretching.
This effect is stronger for flexible chains, which don't have the additional effect of angular constraints.

We further show that deviations from rest positions lead directly to
potential energy being stored in the system in each case.
The potential energy is calculated by summing $U_\mathrm{LJ}$, $U_\mathrm{FENE}$
and $U_\mathrm{bend}$ over every interaction in the minimized configuration,
and is given as a function of chain length and stiffness in Figure~\ref{fig:maps}d.
The main effect is that increasing chain length introduces more FENE bonds into the system,
whose individual deviations of $\sim0.02\sigma$ (Figure~\ref{fig:maps}b)
contribute significantly to increasing stored potential energy.
There is an additional effect whereby stored potential energy
increases with increasing stiffness.
This has contributions from LJ and FENE, in line with their deviations shown in Figures~\ref{fig:maps}a-b, and also from bending potentials.
Although the triplet configurations increasingly conform to their rest positions with increasing stiffness,
remaining deviations become progressively more costly as $\kbend$ increases, leading to a contribution to the stored potenital energy.

\subsection{Overview of structural changes}
The evolution of density and internal stresses
(as parameterized by deviations from interaction energy minima)
are strong functions of the bending stiffness and chain length.
Upon increasing the stiffness, all angles between adjacent bonds
tend to approach the minimum of the bending potential at $\theta_{0}$.
However, this effect competes with the tendency of neighboring beads to stay close to the minima of LJ and FENE interactions.
This competition leads to an increase of potential energy due to pairs of beads drifting subtly away from LJ and FENE minima. 
The effect of increasing the chain length is to insert more FENE bonds into the system.
Since these are typically $\sim0.02\sigma$ from their resting positions,
this leads to a sharp increase in stored potential energy.
We have checked that increasing chain length further above $L=50$ does not bring any further evolution,
and we can safely conclude that chain length has a non-negligible effect only for $L<50$.

\section{Vibrational density of states}

We next investigate the connection between chain length, angular potential and the VDOS.
Importantly, changes in the spectrum due to increasing the angular potential are not only related to the associated increase in angular forces,
but also to the structural changes that arise as discussed above. 
In what follows, we first outline our formalism for obtaining the VDOS,
including a description of how we decompose it into various contributions.
We then give an overview of the generic features of the VDOS of polymeric glasses,
before focussing specifically on the behavior with respect to chain length and bending stiffness.

\subsection{Formalism for obtaining the VDOS}

Since we prepared the glasses well below $T_g^*$,
we can ignore any effects of thermal noise and hence work in the harmonic approximation,
where the displacements of the system around energy minima are small.
The equation of motion can therefore be written with the Hessian $\textbf{H}$ of our system:
        \begin{equation}\label{eq:NewtonsEquation}
        m \,\ddot{\textbf{u}} \,=\, - \textbf{H} \, \textbf{u} \text{.}
        \end{equation}          
Here $m$ is the mass of the constituent beads of our polymer chains (which we take to be uniform)
and $\textbf{u}$ is the displacement field.
We can convert this equation into an eigenvalue problem by performing a Fourier transform,
which gives:
        \begin{equation}\label{eq:EigenvalueProblem}
        m \,\omega^2\,\hat{u} \,=\, \textbf{H} \, \hat{u} \text{,}
        \end{equation}
where $\omega$ are the eigenfrequencies of our system and
$\hat{u}$ are the eigenvectors (displacement fields).
Arranging the eigenfrequencies in a normalized histogram
gives us the vibrational density of states (VDOS) of our system.
To obtain $\omega$, we first need the explicit expressions for the elements of the Hessian,
after which we can solve Equation~\eqref{eq:EigenvalueProblem} numerically.
The elements of the Hessian are defined as second derivatives of the potential energy of the system:
\begin{subequations}
\begin{equation}\label{eq:HessianDef}
        H_{n m}^{a b} \,=\, \frac{\partial^2 \mathcal{U}(z)}{\partial r_n^a\,\partial r_m^b} \text{,}
        \end{equation}    
\begin{equation}
\begin{split}   
\frac{\partial^2 \mathcal{U}(z)}{\partial x\,\partial y}\,&=\, \frac{\partial^2 \mathcal{U}(z)}{\partial z^2} \frac{\partial z}{\partial x} \frac{\partial z}{\partial y} \,+\, \frac{\partial \mathcal{U}(z)}{\partial z} \frac{\partial^2 z}{\partial x\,\partial y}\,\\
&=\, c\, \frac{\partial z}{\partial x} \frac{\partial z}{\partial y} \,+\, t\, \frac{\partial^2 z}{\partial x\,\partial y} \text{.}
\end{split}
\end{equation}
\end{subequations}
Here, $\mathcal{U}$ represents the overall potential,
consisting of the sum of $U_\mathrm{LJ}$, $U_\mathrm{FENE}$ and $U_\mathrm{bend}$,
$z$ is a generic argument,
and $a$ and $b$ label the Cartesian components.
As one can see, the entries in the Hessian consist of two parts:
one proportional to the spring constant $c$ between two beads,
and another one proportional to the tension $t$ (precise definitions are given in Appendix A).
The latter contribution vanishes if all bonds are at their energy minimum at the same time.
In reality this would require perfect crystallisation of the system, which is often not possible or would take a very long time.
In this case, crystallisation is inhibited even for fully flexible chains by the disparity in rest positions of LJ and FENE interactions. 
        
Another source of tension is thermal noise,
though this is not addressed in the present work since we work sufficiently below $T_g^*$.
The main source of tension terms in our simulations is thus the angular potential
and its competition with the LJ and FENE potentials, as discussed above.
This combination of potentials creates two competing effects:
the stronger angular potential forces all angles closer to the rest angle $\theta_0$,
but also increases the strength of the tension for a given deviation,
and generates additional tensions due to increased deviations from the LJ and FENE minima.
Including them in the Hessian, we can now solve Equation~\eqref{eq:EigenvalueProblem} and get the eigenvalues $\omega$
and displacement fields $\hat{u}(\omega)$.
The units of $\omega$ are $\sqrt{\varepsilon_\mathrm{LJ}/m \sigma^2}$.
        
        \begin{figure}
                \centering
                \includegraphics[trim=0cm 6.8cm 0cm 0cm, clip=true,width=0.45\textwidth]{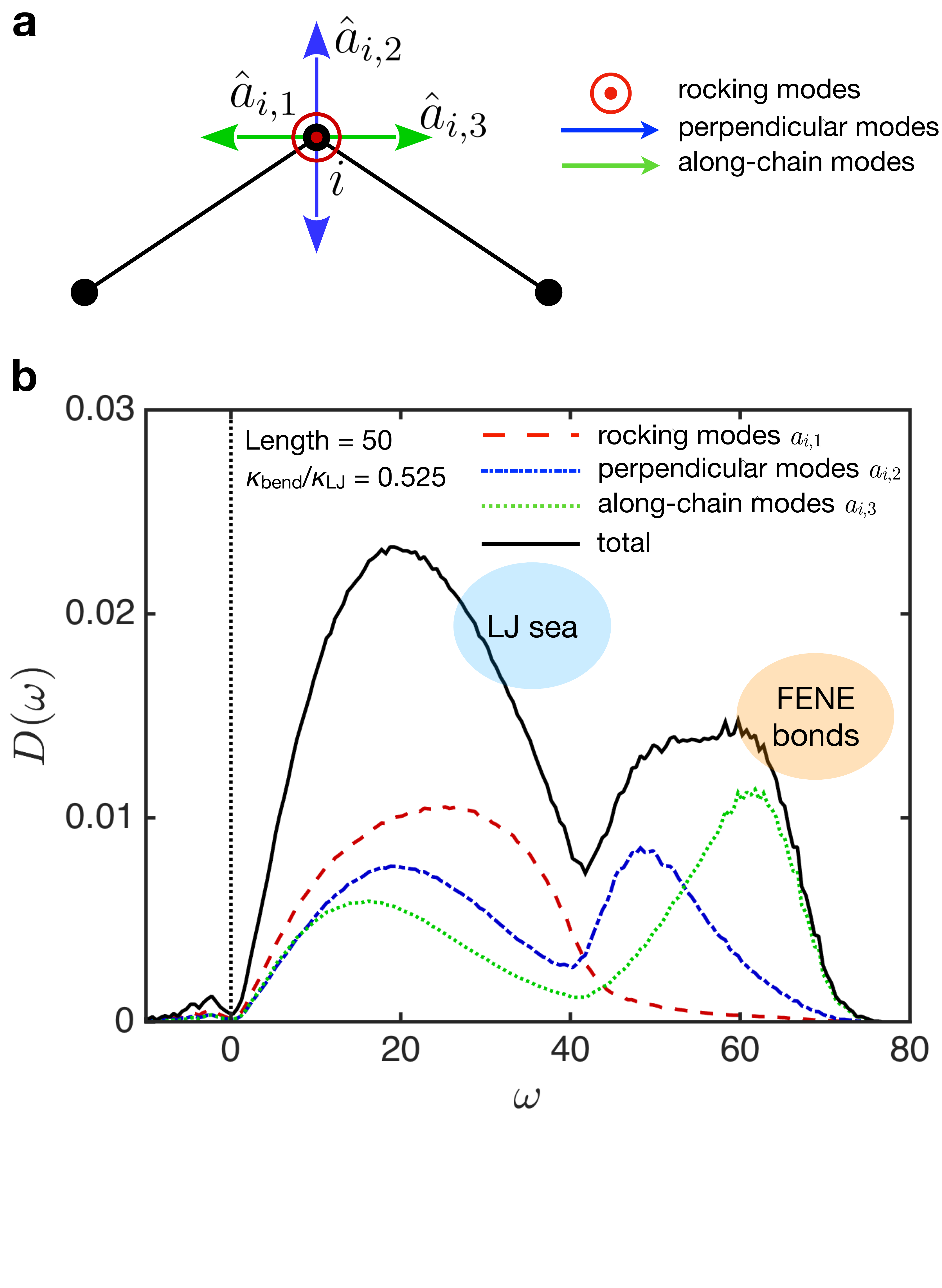}
                \caption{
                (a) Important motions of bead $i$ along the polymer chains. $\hat{a}_{i,1}$, red circle denotes the motion perpendicular to the plane spanned by the two bonds (black lines) also referred to as rocking motion. The other two planar motions are perpendicular ($\hat{a}_{i,2}$, blue arrows) and along the chain ($\hat{a}_{i,3}$, green arrows);
                (b) Example VDOS for chain length $L=50$ and stiffness $\kbend=0.525$, displaying a lower-frequency band corresponding to collective LJ-dominated motions, and a higher-frequency band corresponding to skeletal motions, which include FENE bonds. The dashed line (red) represents rocking motions, the dashed-dotted (blue) line represents perpendicular skeletal motions, while the dotted line (green) represents along-chain skeletal motions.        
                }
                \label{fig:motions}
        \end{figure}

A particularly instructive quantity is the ratio at which different vibration patterns contribute at certain frequencies~\cite{damart2017theory}.
We are especially interested in the internal-coordinate directions shown in Figure~\ref{fig:ChainMotion}.
These motions correspond to out-of-plane ``rocking'' motions (red, $\hat{a}_{i,1}$),
perpendicular (to the chain) motions that remain in the plane of the chain (blue, $\hat{a}_{i,2}$),
and ``along-chain'' motions (green, $\hat{a}_{i,3}$).
To obtain each of these contributions separately,
we project the displacement vector $\hat{u}_i(\omega)$ of each bead
onto the orthogonal basis formed by the three unit vectors
$(\hat{a}_{i,1}, \hat{a}_{i,2}, \hat{a}_{i,3})$
(see Figure~\ref{fig:motions}),
generating a new representation $\hat{v}_i(\omega)$:
        \begin{equation}\label{eq:BaseChange}
        \begin{gathered}
        \hat{v}_i(\omega) \,=\, \begin{pmatrix}
        \hat{u}_i(\omega) \cdot \hat{a}_{i,1}\\
        \hat{u}_i(\omega) \cdot \hat{a}_{i,2}\\
        \hat{u}_i(\omega) \cdot \hat{a}_{i,3}
        \end{pmatrix} = \begin{pmatrix}
        \hat{v}_{i,1}\\
        \hat{v}_{i,2}\\
    \hat{v}_{i,3}
        \end{pmatrix}\\
        \mathcal{X}_j(\omega) \,=\, \sum\limits_{i=1}^N \hat{v}_{i,j}^2 \qquad j = \{1,2,3\}
        \end{gathered}
        \end{equation}
$\mathcal{X}_j(\omega)$ is a weight function which measures the contribution of each of the three different motions discussed above,
from which we get a partial VDOS showing the contribution of each of these three motions to the full VDOS.
This decomposition provides insights into the dynamics of the chains at different frequencies in the spectrum.

\subsection{The vibrational spectrum: results and interpretation}
        \begin{figure*}
        \includegraphics[trim=0mm 22mm 0mm 0mm, clip=true,width = 0.98\textwidth]{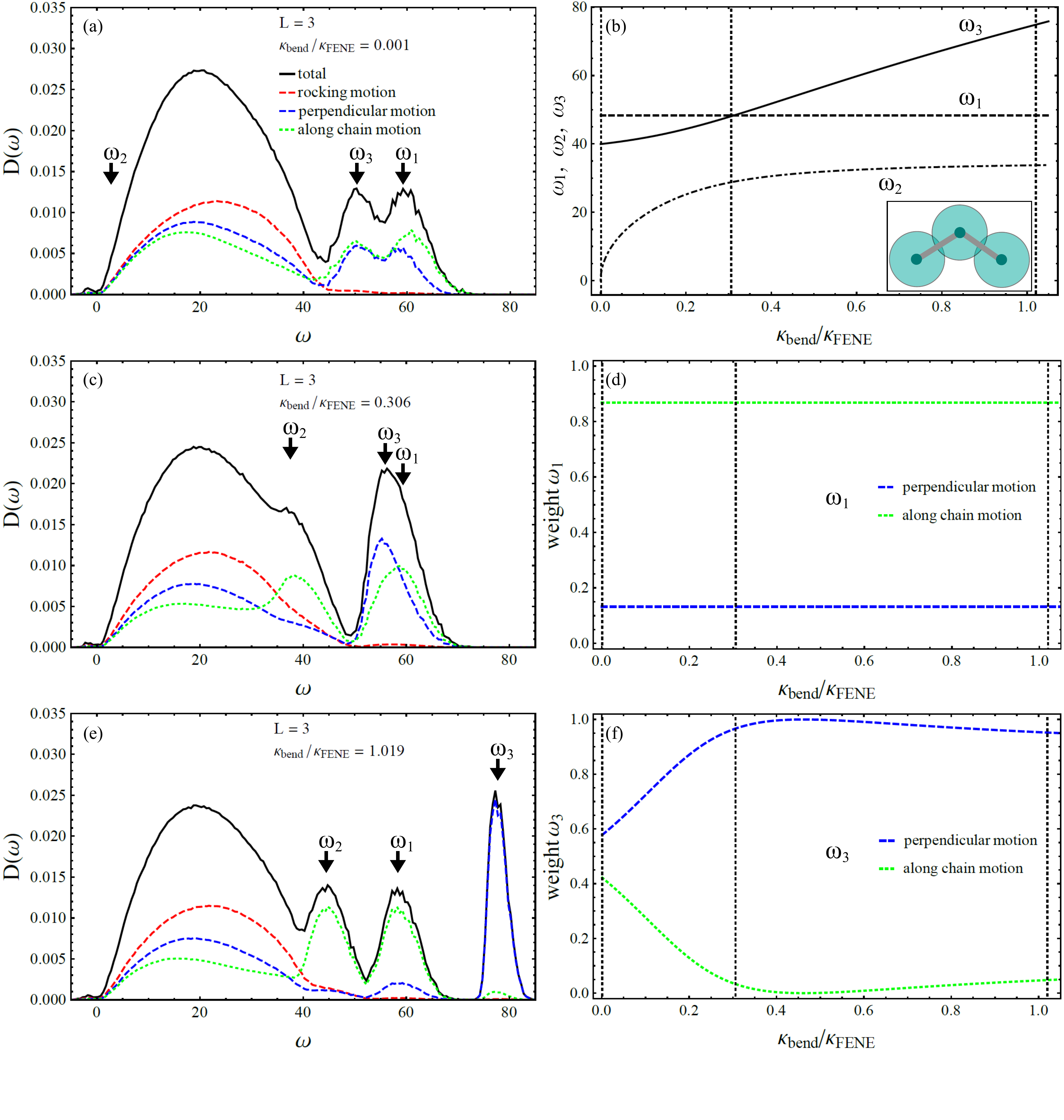}
        \caption{
        VDOS for $L=3$ with
        (a) $\kbend=0.001$;
        (c) $\kbend=0.306$;
        (e) $\kbend=1.019$.
        (b) Analytical solution for the three non-zero eigenfrequencies of a chain with $L=3$ (pictured in Inset) as a function of $\kbend$ (see Appendix~\ref{app:l3}).
        (d,f) Contribution weights for along-chain and perpedicular motion for $\omega_1$ and $\omega_3$ from analytical model. The weights of $\omega_2$ are inverted compared to $\omega_3$.
        Vertical dashed lines in (b,d,f) indicate sample values of $\kfbend$ for which the VDOS are shown in (a,c,e).
        As we can, see the qualitative behavior of our simulated systems with $L=3$ is well-captured by the analytical model, both in terms of frequency and motion weight evolution. The frequencies measured by simulation differ slightly in magnitude due to the large number of LJ interactions that collectively push the bond energy, and therefore frequency, to higher values.}
                \label{fig:dos}
        \end{figure*}

\subsubsection{Collective Lennard-Jones `sea' and higher frequency skeletal modes}
        An example VDOS is given in Figure~\ref{fig:motions},
        for chain length $L=50$ and stiffness $\kbend=0.525$, while in Figure~\ref{fig:dos_split} we present the
        VDOS across the full range of stiffnesses and chain lengths.
        A common feature of this work is
        a distinct splitting of the VDOS into low and high frequency parts,
        particularly evident for low bending stiffness.
        The low frequency part occupies the interval $\omega=[0,\sim40$],
        while the high frequency part extends up to $\omega\approx 70$ in most cases,
        and up to $\omega\approx 100$ when $\kfbend\gtrsim1$.
        The low and high frequency bands are separated by a trough,
        whose depth and precise location in $\omega$ is subtly dependent on $\kbend$.
        This generic splitting of the VDOS into two bands was shown previously by~\citet{jain2004influence},
        who considered fully flexible chains only. 
        Moreover, experimental works in polymerisation have demonstrated that during periods of chain growth a single peak in the Raman intensity transforms into two peaks~\cite{caponi2009raman,caponi2011effect}. These may be related to the two distinct bands predicted here, though a quantitative link between the Raman spectra and the VDOS reported here remains challenging and is the subject of ongoing work.
        We expect, furthermore, that imposing pressures greater than zero will increase the vibrational energy of both bands, thus shifting the spectrum to higher frequencies as observed experimentally~\cite{hong2008pressure}.
        By considering the relative prefactors of the LJ and FENE potentials,
        we find it instructive to interpret the low frequency part as a Lennard-Jones `sea',
        that comprises weak but ubiquitous inter-chain LJ interactions,
        while the high frequency part represents FENE bonds 
        that are fewer in number and follow specific paths along chain backbones.
        Within this picture, the contributions to the VDOS coming from bending interactions are highly sensitive to $\kbend$.
        In particular, when $\kbend$ is small,
        we expect bending interactions to contribute frequencies comparable to, or even lower than, the LJ interactions.
        By contrast, when $\kfbend\to1$ we expect the bending interactions to contribute frequencies comparable to the FENE interactions.
        We anticipate a redistribution, therefore, of the bending contributions from the low to the high frequency band as $\kbend$ is increased.

By analysing the motion patterns with respect to the geometry outlined in Figure~\ref{fig:ChainMotion},
we can see in Figure~\ref{fig:motions} that for flexible chains out-of-plane (rocking) motions are predominantly apparent in the low frequency peak,
while the more energetic modes mainly contain motions in the plane (skeletal vibrations).
This is consistent with the low frequency band being mainly due to LJ interactions,
since they are the only interactions in the system which are not contained in the plane defined by two adjacent bonds.
Consequently, the high frequency part is mostly caused by the FENE bonds,
which point along the chain backbones.

For high bending stiffness we can see a separation between along chain and perpendicular motion,
of which the latter one occupies the high frequency part.
Hence, the three-body bending interaction is mostly associated with perpendicular motion,
whereas the FENE interaction is rather associated with along-chain motions.
This makes sense as our chains have a rest angle $\phi_0 > \pi/2$, meaning the FENE bonds point mostly along the chain direction.

For longer chains we see rocking motion arising at higher frequencies.
The reason for this is that perpendicular in-plane motion caused by one triplet in the chain causes out-of-plane motion from the perspective of neighboring triplets,
since they most likely do not lie in the same plane (as would be the case for a completely flat chain).
As our chains are freely rotating, having a completely flat chain is very unlikely,
which explains why strong bending interaction causes rocking motion at high frequencies.

        \begin{figure}
        \includegraphics[width = 0.45\textwidth]{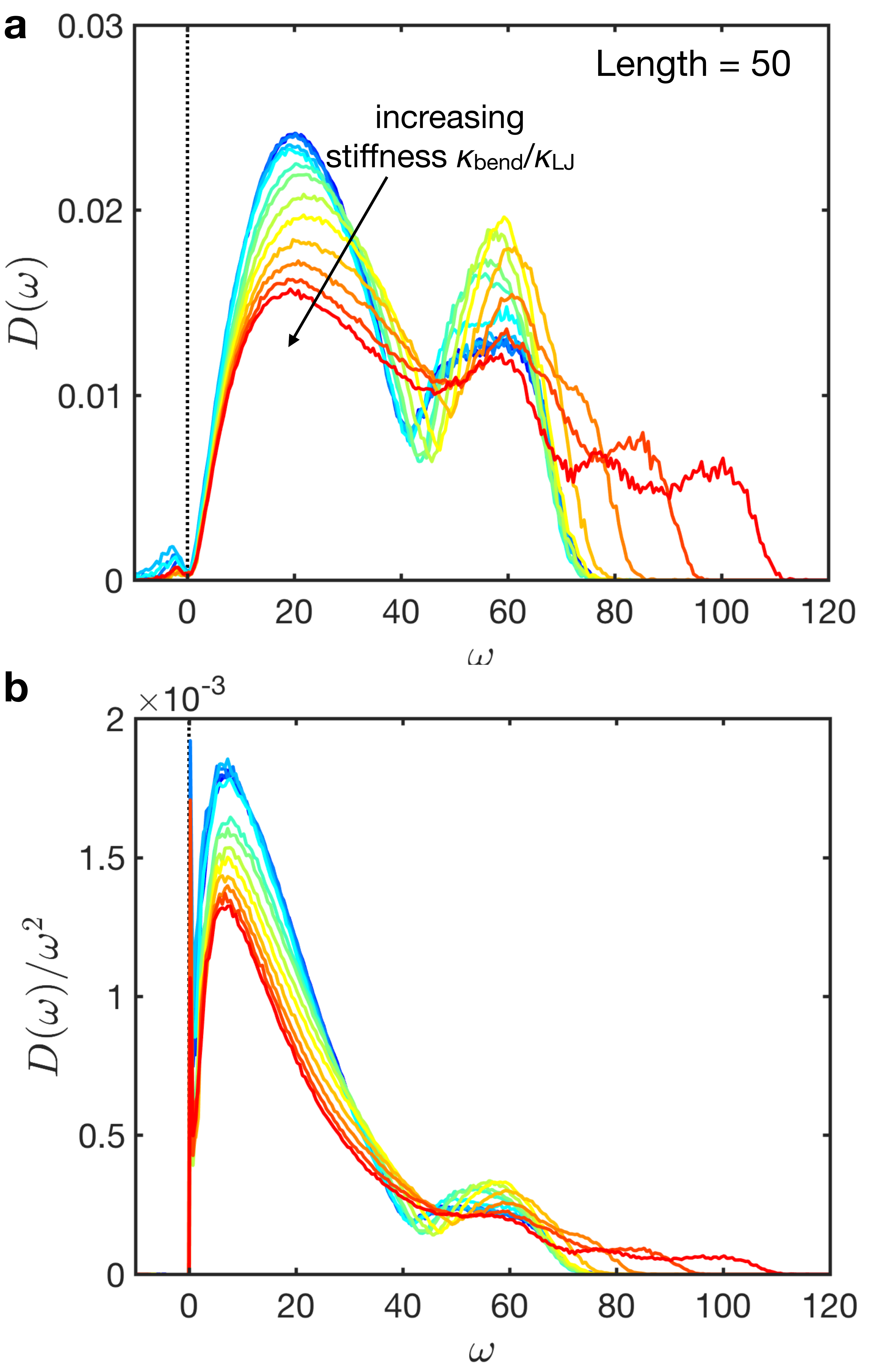}
        \caption{
        (a) Vibrational density of states for $L=50$ upon increasing the stiffness, $\kbend$. The arrow points in the direction of increasing bending stiffness;
        (b) Same data as (a) with the VDOS $D(\omega)$ rescaled by $\omega^2$ to highlight the behavior of the Boson peak.}
                \label{fig:dos_stiffness}
        \end{figure}

\subsubsection{Short chain behavior}
        
For chains with $L=3$, we use the analytical form of the eigenvalues, including both bending and stretching interactions,
to provide insights into the behavior of the VDOS measured in the simulation. The analytical derivation is reported with full details in Appendix~\ref{app:l3}, with the expressions for the eigenvalues given in Equation~\eqref{L3_eigenvalues}.
We present both the analytical and numerical results in Figure~\ref{fig:dos}.

The analysis predicts three non-zero eigenmodes, whose characteristic eigenfrequencies as a function of $\kbend$ are given in Figure~\ref{fig:dos}b.
The splitting of their associated motions into along-chain and perpendicular components are given in Figures~\ref{fig:dos}d,f. Since $L=3$ chains are planar, rocking motions are not part of this analysis. We have seen that rocking motions contribute mostly to the lower LJ-sea band and not so much to the high-frequency skeletal band which is our main focus here.

In Figure~\ref{fig:dos}a,c,e we can see the VDOS for three different values of $\kfbend$ as determined from the simulation at $L=3$. Aside from the LJ-peak (which is not part of the analytical model), we can see that the VDOS follows closely the analytical prediction of the frequencies.
In particular, the peak $\omega_1$ remains dominated by along-chain motions and also remains rather independent of $\kbend$ throughout.
$\omega_2$ is initially at zero (meaning that it is a soft mode) but becomes stiffer and moves to the right as $\kbend$ is increased.
Meanwhile, $\omega_3$ is initially slightly lower than $\omega_1$ but progressively increases also,
eventually crossing over and becoming the higher of the three eigenvalues and at the same time becomes dominated by perpendicular motion.

While the analytical model predicts delta peaks at eigenvalues $\omega_1$, $\omega_2$, $\omega_3$, in practice the peaks are broadened due to the distribution of rest angles, even for very high bending stiffness.
At around $\kfbend=0.3$ the two peaks $\omega_1$ and $\omega_3$ start to overlap and to merge into a single high peak. By looking at the evolution of eigenfrequencies and the associated motions for bending stiffness larger than$\kfbend=0.3$ we can also verify the interpretation that high frequency modes $\omega_3$ are dominated by perpendicular motions for high $\kfbend$, whereas the mode that only depends on the backbone interaction $\omega_1$ is dominated by along-chain motions independently of $\kfbend$. 
The mode $\omega_2$ has weights for along-chain and perpendicular motion which are the specular opposite to the $\omega_3$ case and hence presents a growing along-chain character upon increasing $\kfbend$.

VDOS for additional values of $\kfbend$ are shown in Figure~\ref{fig:dos_split}. We can clearly see the continuous shift of modes according to the analytical result given in Equation~\eqref{L3_eigenvalues}, reflected also in the motion pattern associated with those modes as shown in Figure~\ref{fig:dos}.
        
\subsubsection{Dependence on bending stiffness for $L>3$}
\begin{figure*}
        \includegraphics[width = 1\textwidth]{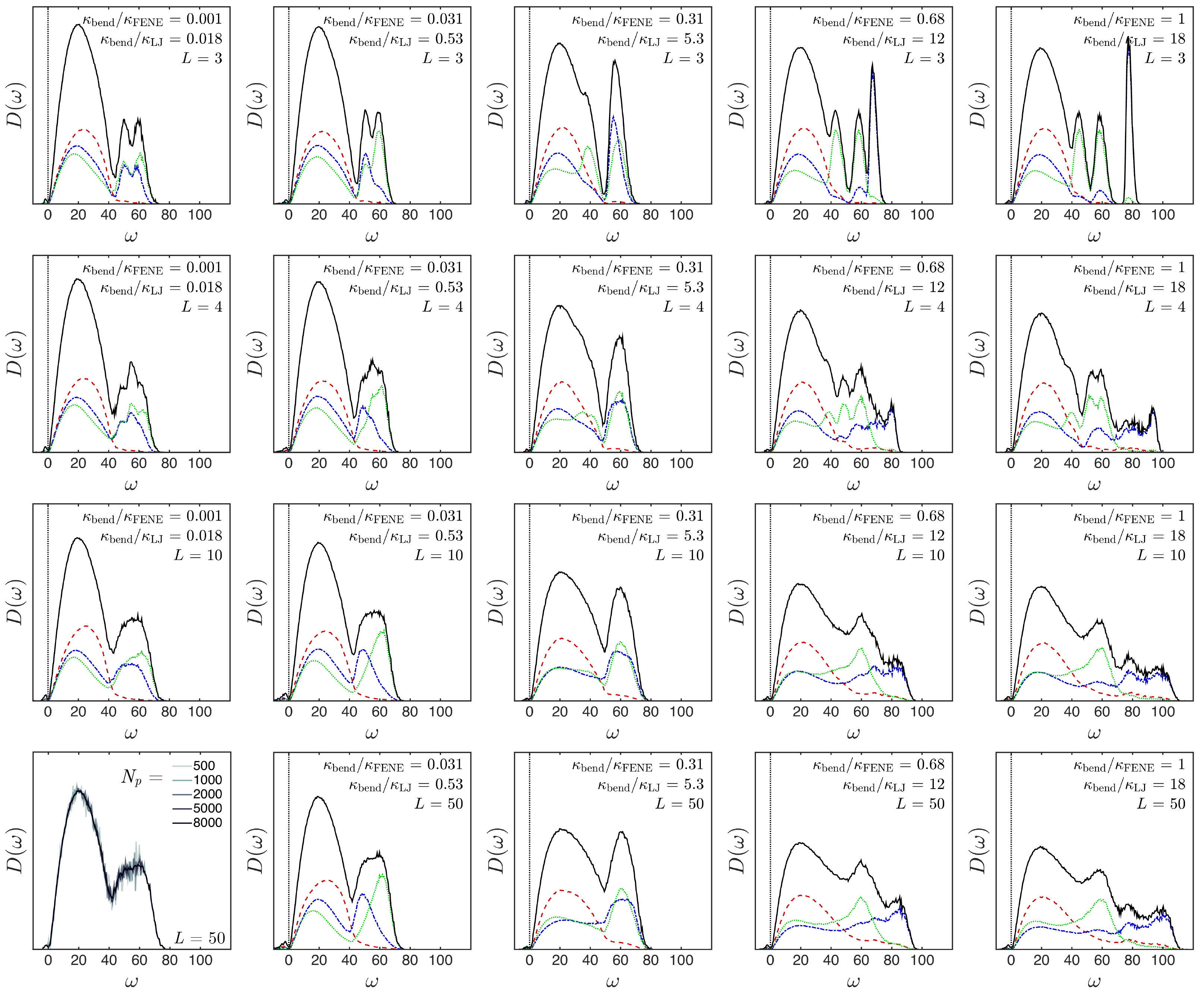}
        \caption{Vibrational density of states for a range of parameters explored in this work. Shown are the overall result in black, as well as the decomposition into rocking (red), perpendicular (blue) and along-chain (green) motions as we defined in figure~\ref{fig:motions}. Given in the legend of each panel are the chain length and stiffness. Results for various system sizes are shown in the bottom left panel, demonstrating that the features discussed here are independent of $N_p$.}\label{fig:dos_split}
        \end{figure*}

        For longer chains we can see the same general features as for $L=3$, Figure~\ref{fig:dos_stiffness} and Figure~\ref{fig:dos_split}.
        Higher bending stiffness leads to redistribution of modes from the lower part of the FENE regime towards higher frequencies with an overlap happening at $\kfbend\approx 0.3$,
        where they form a single peak.
        Above that value the bending interaction shifts modes associated with perpendicular motions towards higher frequencies, while the modes associated with along-chain motion stay relatively unchanged.
        
The strong bending interaction also causes high frequency out-of-plane motions to appear, as discussed above.
The gap between the LJ-sea and the FENE band is filled by modes in the same way as the third peak arises for $L=3$. We can relate the peak at $\omega = 60$ to $\omega_1$ from the $L=3$ model system by looking at a second toy model, the freely rotating chain with constant bond angle but no bending potential, as described in Appendix C, Equations~\eqref{eq:HessianChain}-\eqref{eq:motion_weights}. The VDOS is given by Equation~\eqref{roots_regular}, and is U-shaped spectrum with two divergent peaks at the van-Hove singularities (similar to the textbook example of completely straight linear chains with all angles at $180^{o}$). Since $\omega_1$ does not change with the bending stiffness it is natural that a remnant of this peak related to this frequency appears stationary for all bending stiffness.
The lower peak, however, would correspond to $\omega_3$, which depends heavily on the bending stiffness and can therefore not be fully captured by a toy model without bending interaction.
        
We further studied the so-called Boson peak, defined as the excess of low-frequency modes above the Debye $\sim \omega^{2}$ law, which is a paradigmatic and defining feature of glasses. In Figure~\ref{fig:dos_stiffness}b the VDOS normalized by the Debye law is shown, and it is evident that increasing the bending stiffness causes a significant decrease of the Boson peak. This is due to the fact that, since the VDOS is a normalized distribution, if vibration modes are shifted to high frequency due to the stiffening of skeletal vibrations involving bending, then necessarily the density of modes has to decrease in lower-frequency parts of the spectrum. From the point of view of mechanical response, a decrease of the Boson peak is linked with a decrease of the nonaffine component of elastic~\cite{lemaitre2006sum,zaccone2011approximate} and viscoelastic~\cite{rico2017viscoelastic} response which contributes negatively to the shear modulus.

        Hence, increasing the bending stiffness has a twofold effect on the elasticity: it increases the affine part of the shear modulus (which is a positive contribution to rigidity) by increasing the stiffness constant, and it decreases the Boson peak and therefore decreases (in absolute value) the nonaffine part of the shear modulus (which is a negative contribution to rigidity), as explained in previous work~\cite{lemaitre2006sum,zaccone2011approximate}. This, however, does not account for the structural effect brought about by increasing stiffness, which leads to volumetric expansion and, under certain conditions (as discussed by~\citet{ness2017nonmonotonic}) may lead to a decrease of shear modulus upon increasing bending stiffness, thus giving rise to a non-monotonic dependence of the shear modulus on chain stiffness. 
        Clearly this mechanism by which the Boson peak is changed in polymer glasses, is very different from other mechanisms discussed in the literature for small-molecule or atomic glasses~\cite{mizuno2014acoustic,milkus2016local}.
        
        To summarize, we can separate the spectrum into four distinct parts:\\
        i) the lowest-frequency band of LJ-sea which gets lowered as, with higher bending stiffness, more modes are shifted to higher frequencies;\\
        ii) a stationary peak around $\omega = 60$ strongly associated with along-chain motion/vibration and with characteristic frequency $\omega_1$;\\
        iii) modes associated with vibrations perpendicular to the chain axis, that resemble the behavior of $\omega_3$ and the frequencies of which diverge with the bending stiffness;\\
        iv) modes associated with along-chain vibrations that resemble the behavior of $\omega_2$ filling the regime between the LJ-sea and the 
        $\omega_1$-peak.\\

        \subsubsection{Dependence on length $L$}

        By increasing the chain length we introduce more high-energy FENE bonds into the system, which leads to a shift or redistribution of modes from the low to the high frequency band, causing a lowering of the LJ peak (Figure~\ref{fig:dos_split}). This is particularly evident for fully flexible chains. Additionally, the number of possible polymer configurations increases drastically with chain length. In the VDOS this leads to a loss of distinct features (i.e. sharp peaks become broadened), especially in the high frequency band. The effect is best visible for fully flexible short chains of $L=3,4,5$, whereas the difference between $L=10$ and $L=50$ is only marginal.
The total number of additional FENE bonds per chain $n$ decreases as the chains become longer $n/N = \frac{L - 1}{L} \,,\, \Delta n/N = \frac{1}{L (L + 1)}$
and therefore the change in total bond energy becomes smaller.
As such, the averaged spectrum of sufficiently long ($L > 5$) chains already approximates the spectrum of an infinitely long chain quite well.
A sample of chains with $L = 10$ consequently shows the same shape as for $L = 50$ (or even $L = 1000$).
We thus limited our analysis to systems with those lengths, as there is no new physics to see in the spectrum of chains longer than $L = 50$.
        The distributions of motion patterns do not change much with the chain length, aside from the adjusting to the overall shape of the spectrum described above.
        An analytical derivation for fully flexible linear chains with $L=2,3,4,5$, with stretching interactions, is reported with full details in Appendix C.
        
        \section{Concluding remarks}
        We presented a systematic analysis and interpretation of the structure,
        internal stresses and vibrational spectra of glassy polymers from coarse-grained simulations,
        based on the Kremer-Grest bead-spring model with an energy minimum for angular bending interaction.        
        Varying the angular stiffness and the chain length leads to rich phenomenology:
        an increase in both these parameters causes a build-up of internal stresses
        due to the competition between bending and stretching degrees of freedom,
        both of which want to minimize their energy at the same time.
        This leads to increased deviations from the minima of LJ and FENE interactions, an effect amplified for longer chains.

        For flexible chains with $\kfbend<0.3$ there are two bands in the VDOS spectra,
        corresponding to LJ-dominated interactions at low frequency (the LJ-sea),
        and to skeletal modes dominated by FENE bonds at high frequency (the high-frequency skeletal band).
        For chains with higher bending stiffness this separation breaks down as modes associated with angular interactions appear, filling the gap between the two bands and creating additional high frequency modes.
        The latter of these are mostly made up from vibrations perpendicular to the chain,
        while the gap is filled by along-chain motions,
        creating a new separation between vibrational regimes in the spectrum.
        
        For short chains, the spectra feature sharp peaks whose behavior as a function of stiffness correspond very well with our analytical prediction.
        For longer chains this structure loses its sharp features and tends to a more continuous spectrum in which the various peaks are broadened by the averaging over many different chain conformations.
        
        This framework and concepts can be applied in future work to molecular and atomistic simulations of realistic materials~\cite{rutledge2016polyether,rutledge2015mechanics},
        possibly in combination with Kernel Polynomial Methods~\cite{beltukov2016KPM} which can greatly speed up the evaluation of the VDOS using the Hessian as input. 
        We anticipate that the generic features of the VDOS predicted in this work will be robust to the introduction of more specific chemical interactions (including those that break the isotropic interaction symmetry), since the features of the vibrational spectrum are related essentially to the energy and the relative strength of interactions.
For example, hydrogen bonds (which have typical energy (4-13kJ/mol) an order of magnitude less than covalent C-C bonds (346kJ/mol)) would be expected to add to the part of the spectrum that is already dominated by the LJ sea. The same can be stated about stacking interactions, which have typical energies of 8-12 kJ/mol.
        
        Finally, our results may open up the possibility of quantitatively linking the Raman and Brillouin spectra of glassy polymers with their viscoelastic response, since the VDOS is a key input to calculate viscoelastic moduli within recent developments in the nonaffine linear response of amorphous solids~\cite{rico2017viscoelastic,damart2017theory}.

        \section{Acknowledgements}

We had useful discussions with Tim Sirk and Robert Elder. C.N. acknowledges the Maudslay-Butler Research Fellowship at Pembroke College, Cambridge for financial support.
V.V.P. and A.Z. acknowledge financial support from the U.S. Army ITC-Atlantic and the U.S.
Army Research Laboratory under cooperative Agreement No.
W911NF-16-2-0091.
The research leading to these results has also received funding from the European Research Council under the European Union's H2020 Grant Agreement no. 636820.
\appendix{
        \section{Explicit form of the Hessian}\label{app:A}

        Here we show the explicit form of the entries to the Hessian for each potential. First we recall the general form: 
        \begin{subequations}
\begin{equation}\label{eq:HessianDefAppendix}
        H_{n m}^{a b} \,=\, \frac{\partial^2 \mathcal{U}(\textbf{r})}{\partial r_n^a\,\partial r_m^b},
        \end{equation}    \\
where $a$ and $b$ label the Cartesian components, and the entries are given, for a generic argument $z$, as:
\begin{equation}
\begin{split}   
\frac{\partial^2 \mathcal{U}(z)}{\partial x\,\partial y}\,&=\, \frac{\partial^2 \mathcal{U}(z)}{\partial z^2} \frac{\partial z}{\partial x} \frac{\partial z}{\partial y} \,+\, \frac{\partial \mathcal{U}(z)}{\partial z} \frac{\partial^2 z}{\partial x\,\partial y}\,\\
&=\, c\, \frac{\partial z}{\partial x} \frac{\partial z}{\partial y} \,+\, t\, \frac{\partial^2 z}{\partial x\,\partial y}.
\end{split}
\end{equation}
\end{subequations}
        For the two central-force potentials (FENE und Lennard-Jones) we have $z = \lvert \textbf{r}_{j}-\textbf{r}_{i} \rvert = r_{i j}$. It should be noted that all derivatives are evaluated at the actual configuration from the simulation. For central forces we get:
        \begin{equation}\label{eq:HessianCentral}
        \begin{gathered}
        H_{n m}^{a b} \,=\, \frac{\partial^2 \mathcal{U}(r_{i j})}{\partial r_n^a\,\partial r_m^b} = c_{i j} \frac{\partial r_{i j}}{\partial r_n^a} \frac{\partial r_{i j}}{\partial r_m^b} \,+\, t_{i j} \frac{\partial^2 r_{i j}}{\partial r_n^a\,\partial r_m^b},\\
        c_{i j} \,=\, \frac{\partial^2 \mathcal{U}(r_{i j})}{\partial r_{i j}^2}\quad;\quad t_{i j} \,=\, \frac{\partial \mathcal{U}(r_{i j})}{\partial r_{i j}},\\
        \frac{\partial r_{i j}}{\partial r_n^a}\,=\, \left( \delta_{n j} - \delta_{n i} \right) \hat{n}_{i j}^a,\\
        \frac{\partial^2 r_{i j}}{\partial r_n^a\,\partial r_m^b}\,=\, \frac{1}{r_{i j}}\left( \delta_{n j} - \delta_{n i} \right)\left( \delta_{m j} - \delta_{m i} \right) (1 - \hat{n}_{i j}^a \hat{n}_{i j}^b).\\
        \end{gathered}
        \end{equation}
        Here $\hat{n}_{i j} = \textbf{r}_{i j}/r_{i j}$ denotes the unit bond vector between bead $i$ and $j$. The above expressions are valid both for FENE and Lennard-Jones bonds with the only difference being the stiffness $c_{i j}$ and tension $t_{i j}$ that have to be evaluated depending on the potential.
        For the angular potential we have a slightly different situation as our variable is now the angle between the two bonds $\textbf{r}_{j}-\textbf{r}_{i}$ and $\textbf{r}_{k}-\textbf{r}_{i}$:
        \begin{equation}\label{eq:Angle}
        \begin{gathered}
        z \,=\, \theta_{i j k} \,=\,\arccos  \frac{(\textbf{r}_{j}-\textbf{r}_{i})\cdot(\textbf{r}_{k}-\textbf{r}_{i})}{\lvert \textbf{r}_{j}-\textbf{r}_{i} \rvert\,\, \lvert \textbf{r}_{k}-\textbf{r}_{i} \rvert } = \arccos A_{i j k}.
        \end{gathered}
        \end{equation}
        To make the calculation easier we rewrite the first line of \eqref{eq:HessianCentral} to give:
        
        \begin{widetext}
        \begin{equation}\label{eq:HessianDefAngle}
        \begin{gathered}
    H_{n m}^{a b} \,=\, \frac{\partial^2 \mathcal{U}(\theta_{i j k})}{\partial r_n^a\,\partial r_m^b} = \tilde{c}_{i j k} \frac{\partial \theta_{i j k}}{\partial r_n^a} \frac{\partial \theta_{i j k}}{\partial r_m^b} \,+\, \tilde{t}_{i j k} \frac{\partial^2 \theta_{i j k}}{\partial r_n^a\,\partial r_m^b}\\
    =\,\frac{\tilde{c}_{i j k}}{\sin^2 \theta_{i j k}} \frac{\partial A_{i j k}}{\partial r_n^a} \frac{\partial A_{i j k}}{\partial r_m^b} \,-\, \frac{\tilde{t}_{i j k}}{\sin \theta_{i j k}} \left( \frac{\tan \theta_{i j k}}{\sin \theta_{i j k}} \frac{\partial A_{i j k}}{\partial r_n^a} \frac{\partial A_{i j k}}{\partial r_m^b} + \frac{1}{\sin \theta_{i j k}} \frac{\partial^2 A_{i j k}}{\partial r_n^a\,\partial r_m^b} \right),\\
    \frac{\partial A_{i j k}}{\partial r_n^a}\,=\, \frac{1}{r_{i j}} \left( \delta_{n j} - \delta_{n i} \right) \left( \hat{n}_{i k}^a - \hat{n}_{i j}^a \cos \theta_{i j k} \right)+\frac{1}{r_{i k}} \left( \delta_{n k} - \delta_{n i} \right) \left( \hat{n}_{i j}^a - \hat{n}_{i k}^a \cos \theta_{i j k} \right),\\
    \frac{\partial^2 A}{\partial r_n^a\,\partial r_m^b} \,=\, \frac{\delta^n_{j i} \delta^m_{j i}}{r_{i j}^2}\left[ \left( 3\, \hat{n}_{i j}^a\hat{n}_{i j}^b - \delta_{a b} \right) \cos \theta_{i j k} - \left( \hat{n}_{i j}^a\hat{n}_{i k}^b + \hat{n}_{i k}^a\hat{n}_{i j}^b \right) \right]\\
    + \frac{\delta^n_{j i} \delta^m_{k i}}{r_{i j} r_{i k}}\left[ \delta_{a b} + \hat{n}_{i j}^a\hat{n}_{i k}^b \cos \theta_{i j k} - \left( \hat{n}_{i j}^a\hat{n}_{i j}^b + \hat{n}_{i k}^a\hat{n}_{i k}^b \right) \right]\\
    + \frac{\delta^n_{k i} \delta^m_{j i}}{r_{i j} r_{i k}}\left[ \delta_{a b} + \hat{n}_{i k}^a\hat{n}_{i j}^b \cos \theta_{i j k} - \left( \hat{n}_{i j}^a\hat{n}_{i j}^b + \hat{n}_{i k}^a\hat{n}_{i k}^b \right) \right]\\
    + \frac{\delta^n_{k i} \delta^m_{k i}}{r_{i k}^2}\left[ \left( 3\, \hat{n}_{i k}^a\hat{n}_{i k}^b - \delta_{a b} \right) \cos \theta_{i j k} - \left( \hat{n}_{i j}^a\hat{n}_{i k}^b + \hat{n}_{i k}^a\hat{n}_{i j}^b \right) \right].
        \end{gathered}
        \end{equation}
        \end{widetext}
        
        These formulae were derived in a slightly different but substantially equivalent fashion by~\citet{van2006expressions}.

        \section{Analytical solution for triatomic molecule with bending stiffness}
        \label{app:l3}
        We next write down the Hessian for an isolated oligomer with $L=3$ (a triatomic molecule model), accounting for both stretching and bond-bending interactions. As the eigenvalues of the Hessian are invariant under spatial rotations, we can chose the chain lying flat in the x-y plane with beads $P_1 = - r (\varsigma,0)$, $P_2 = r (0, \upsilon)$, $P_3 = r (\varsigma,0)$ and $\varsigma = \sin \theta/2$, $\upsilon = \cos \theta/2$:
        \begin{widetext}        
                \begin{equation}
                \begin{gathered}
                H = \frac{\kappa}{m} \begin{pmatrix}
                \varsigma^2 + \gamma' \upsilon^2 & -(1-\gamma')\,\varsigma\,\upsilon & - \varsigma^2 & (1 - 2 \gamma')\,\varsigma\,\upsilon & - \gamma' \varsigma^2 & \gamma'\,\varsigma\,\upsilon\\
                -(1-\gamma')\,\varsigma\,\upsilon & \varsigma^2 + \gamma' \varsigma^2 & \varsigma\,\upsilon & - \upsilon^2 - 2 \gamma' \varsigma^2 & -\gamma'\,\varsigma\,\upsilon & \gamma' \varsigma^2\\
                - \varsigma^2 &\varsigma\,\upsilon & 2 \varsigma^2 & 0 & -\varsigma^2 & -\varsigma\,\upsilon\\
                (1 - 2 \gamma')\,\varsigma\,\upsilon & - \upsilon^2 - 2 \gamma' \varsigma^2 & 0 & 2 \upsilon^2 + 4 \gamma' \varsigma^2 & -(1 - 2 \gamma')\,\varsigma\,\upsilon & - \upsilon^2 - 2 \gamma' \varsigma^2\\
                - \gamma' \upsilon^2 & -\gamma' \varsigma\,\upsilon & -\varsigma^2 & -(1 - 2 \gamma')\,\varsigma\,\upsilon & \varsigma^2 + \gamma' \upsilon^2 & (1-\gamma')\,\varsigma\,\upsilon\\
                \gamma' \varsigma\,\upsilon & \gamma' \varsigma^2 & -\gamma' \varsigma\,\upsilon & -\upsilon^2 - 2 \gamma' \varsigma^2 & (1-\gamma')\,\varsigma\,\upsilon &\upsilon^2 + \gamma' \varsigma^2
                \end{pmatrix}.\\
                \end{gathered}
                \end{equation}
        \end{widetext}
        Here we used the dimensionless bending stiffness $\gamma' = \gamma/(\kappa\, r^2)$ for bond length $r$, where $\gamma$ (which has units of energy) is the second derivative of the angular bending potential with respect to the angle, while $\kappa$ is the spring constant of the bond for central-force stretching of the bond. The above matrix has three non-zero eigenvalues, leading to the following eigenfrequencies:
        \begin{widetext}
        \begin{equation}\label{L3_eigenvalues}
        \begin{gathered}
        \omega_1^2 = \frac{\kappa}{m}(2 - \cos \theta),\\
        \omega_{2}^2 = \frac{\kappa}{2 m}\left( 2(1+2 \gamma')+(1-2\gamma')\cos \theta - \sqrt{(2(1+2 \gamma')+(1-2\gamma')\cos \theta)^2 - 24 \gamma'} \right),\\
        \omega_{3}^2 = \frac{\kappa}{2 m}\left( 2(1+2 \gamma')+(1-2\gamma')\cos \theta + \sqrt{(2(1+2 \gamma')+(1-2\gamma')\cos \theta)^2 - 24 \gamma'} \right).
        \end{gathered}
        \end{equation}
        \end{widetext}
        As we can see, one eigenfrequency $\omega_1$ is independent of the bending stiffness, while $\omega_{2}$ shows a convergent behavior against $\omega \rightarrow \sqrt{3\kappa/m/(2 - \cos \theta)}$ and $\omega_3$ diverges like $\sim \sqrt{\gamma'}$.

        \section{Analytical solutions for fully-flexible chains}
In this Appendix we consider a toy model for the determination of the skeletal vibration modes of a single polymer chain. The following assumptions are made: (i) the chain is fully flexible (vanishing angular stiffness); (ii) only in-plane motions are considered (rocking or other out-of-plane vibrations are neglected). These assumptions are needed to obtain analytical results. We will start with the simplest case of a zig-zag regular chain with a single fixed value $\theta$ of the angle between two adjacent bonds, and we will subsequently consider the case of a distributed $\theta$.
We consider two variations of this model, first for a zig-zag chain with fixed value of the angle, and subsequently for a uniform (random) distribution of the angle. 
        
        The Hessian has the following block structure:
\begin{widetext}        
        \begin{equation}\label{eq:HessianChain}
        \begin{gathered}
        H =  \begin{pmatrix}
        A_{1 2} & -A_{1 2} & 0 & 0 & 0 & & 0 \\
        -A_{1 2} & A_{1 2} + A_{2 3} & -A_{2 3} & 0 & 0 & \cdots & 0\\
        0 & -A_{2 3} & A_{2 3} + A_{3 4} & -A_{3 4} & 0 & & 0 \\
        0 & 0 & -A_{3 4} & A_{3 4} + A_{4 5} & -A_{4 5} & & \vdots \\
          & \vdots & & & & \ddots & -A_{L-1 L} \\
        0 & 0 & 0 & 0 & \cdots & -A_{L-1 L} & A_{L-1 L}
        \end{pmatrix}\\
        \end{gathered}
        \end{equation}
\end{widetext}  
with blocks given by:
\begin{equation}
A_{i j} = \frac{\kappa}{m} \begin{pmatrix}
        n_{i j}^x n_{i j}^x & n_{i j}^x n_{i j}^y & n_{i j}^x n_{i j}^z \\
        n_{i j}^y n_{i j}^x & n_{i j}^y n_{i j}^y & n_{i j}^y n_{i j}^z \\
        n_{i j}^z n_{i j}^x & n_{i j}^z n_{i j}^y & n_{i j}^z n_{i j}^z \\
        \end{pmatrix},
\end{equation}

        where the unit vector $\mathbf{n}_{i j}$ which goes from bead $i$ to a nearest-neighbor $j$, and spring constant $\kappa=1$. To get the characteristic polynomial $p(\lambda)$, one has to evaluate the determinant $\left\lvert H - \lambda \mathbf{1} \right\rvert$. We can iteratively solve this by using the standard formula for block matrices:
        \begin{equation}
        \begin{gathered}
        \begin{vmatrix}
    A & B \\
    C & D
        \end{vmatrix} = \begin{vmatrix}
        D
        \end{vmatrix} \begin{vmatrix}
        A - B D^{-1} C
        \end{vmatrix}   
        \end{gathered},
        \end{equation}
        where the entries are matrices and the relation for our $3 \times 3$ blocks:
        \begin{equation}
        A \cdot B \cdot A = \frac{\kappa^2}{m^2}\cos^2 \theta_{A B} A,
        \end{equation}
        where $\theta_{A B}$ is the angle between the bonds belonging to $A$ and $B$. After some calculation we get the following recursion formula (omitting the $2 L + 1$ trivial eigenvalues $\lambda = 0$) for the above matrix:
        \begin{equation}\label{recursion}
        \begin{gathered}
        p_0(x) = 1,\\
        p_1(x) = x,\\
        p_n(x) = x\, p_{n-1}(x) - \cos^2 \theta_{n-1 n-2}\, p_{n-2}(x),\\
        \end{gathered}
        \end{equation}
        where $x = \frac{m}{\kappa}\lambda - 2$. The recursive relation for $p_n(x)$ can also be derived on more formal grounds~\cite{kiparissides1996polymerization}.

        Note that $n$ denotes the number of bonds in a chain, and not the number of beads in the chain ($n = L-1$).For arbitrary angles between the bonds it is not possible to describe the roots of this polynomial, except for oligomers (see below). But, if all angles are the same we can bring \eqref{recursion} into the form of the Chebyshev polynomials of the second kind $U_n(x)$ by substituting $\tilde{x} = x/2 \cos \theta$:
        \begin{equation}\label{results_regular}
        \begin{gathered}
        p_n(\tilde{x}) = \cos^n \theta \,\,U_n(\tilde{x}),\\
        \tilde{x} = \frac{x}{2 \cos \theta} = \frac{\frac{m}{\kappa} \lambda - 2}{2 \cos \theta}.\\
        \end{gathered}
        \end{equation}
        \\
        The roots of $U_n(x)$ are $x_k = \cos \left( \frac{k}{n + 1} \pi \right) ; k = 1,...,n$, which gives us the eigenvalues of the linear chain with constant angle as:
        \begin{equation}\label{roots_regular}
        \begin{gathered}
        \lambda_k = \omega_k^2 = \frac{2\kappa}{m}\left( 1 + \cos \theta \, \cos \left[ \frac{k \pi}{n + 1} \right] \right),\\
        D(\omega) = \frac{2}{\pi} \frac{\omega}{\sqrt{\frac{4\kappa^2}{m^2} \cos^2 \theta - (\omega^2 - \frac{2\kappa}{m})^2}}\quad (\theta \neq \pi/2),\\
        \rho(\omega) = \delta\left(\omega - \sqrt{\frac{2\kappa}{m}}\right)\quad (\theta = \pi/2).
        \end{gathered}
        \end{equation}
        This result can also be derived using a different approach which exploits the periodicity of the chain with constant angle, as was done by Kirkwood~\cite{kirkwood1939skeletal}. If we assume that the chain points along the x-axis we can identify the previously introduced along-chain and perpendicular motion as $A$ and $B$ in Eq. 6 of~\citet{kirkwood1939skeletal}. By using the dispersion relation found in this work, we can solve for those two quantities and find the weight functions:
        \begin{equation}\label{eq:motion_weights}
        \begin{gathered}
        X_l(\omega) \,=\, \frac{\lvert A \rvert^2}{\lvert A \rvert^2 + \lvert B \rvert^2}\,=\,\frac{\cos \theta + 1}{\cos \theta}\frac{\omega^2 + \frac{2\omega}{m}(\cos \theta - 1)}{2 \omega^2},\\
        X_t(\omega) \,=\, \frac{\lvert B \rvert^2}{\lvert A \rvert^2 + \lvert B \rvert^2}\,=\,\frac{\cos \theta - 1}{\cos \theta}\frac{\omega^2 - \frac{2\kappa}{m}(\cos \theta + 1)}{2 \omega^2}.
        \end{gathered}
        \end{equation}
        As mentioned before, for the flexible case with distributed angles an analytical solution is not accessible for a chain of arbitrary length. But we can give the eigenvalues in the case of short chains with $L=2,3,4,5$: 
        \begin{widetext}
        \begin{equation}\label{short chains}
        \begin{gathered}
        L = 2: \quad \frac{m}{\kappa} \omega^2 = 2,\\
        L = 3: \quad \frac{m}{\kappa}\omega^2 = 2 \pm \cos \theta_{1},\\
        L = 4: \quad \frac{m}{\kappa}\omega^2 = 2,\,2 \pm \sqrt{\cos^2 \theta_{1}+\cos^2 \theta_{2}},\\
        L = 5: \quad \frac{m}{\kappa}\omega^2 = \pm \frac{1}{\sqrt{2}} \sqrt{\cos^2 \theta_{1}+\cos^2 \theta_{2}+\cos^2 \theta_{3} \pm \sqrt{(\cos^2 \theta_{1}+\cos^2 \theta_{2}+\cos^2 \theta_{3})^2 - 4 \cos^2 \theta_{1} \cos^2 \theta_{3}} }.
        \end{gathered}
        \end{equation}
        \end{widetext}

}


%

\end{document}